\documentclass[pdflatex,sn-mathphys-ed,iicol]{sn-jnl} 

\usepackage{lipsum}
\usepackage{lmodern}
\usepackage{slantsc}
\usepackage{slashed}
\usepackage{mathtools}
\usepackage{siunitx}
\usepackage{xcolor}

\DeclarePairedDelimiterXPP\BigOSI[2]%
  {\mathcal{O}}{(}{)}{}%
  {\SI{#1}{#2}}

  \usepackage{xcolor}
\definecolor{color3}{RGB}{51,51,200}

\hypersetup{
     colorlinks=true,
     linkcolor=color3,
     urlcolor=color3,
     }

\jyear{2026}%

\theoremstyle{thmstyleone}%

\theoremstyle{thmstyletwo}%

\theoremstyle{thmstylethree}%

\begin{document}

\title[]{Minicharged Particles at Accelerators: Progress and Prospects}

\author[1]{\fnm{Marc} \sur{de Montigny}}

\author[2]{\fnm{Pierre-Philippe A.} \sur{Ouimet}}

\author[3]{\fnm{James} \sur{Pinfold}}

\author[3]{\fnm{Ameir} \sur{Shaa}}

\author*[3,4]{\fnm{Michael} \sur{Staelens}}\email{\href{mailto:michael.anthony.staelens@cern.ch}{michael.anthony.staelens@cern.ch}}

\affil[1]{\orgdiv{Facult\'e Saint-Jean}, \orgname{University of Alberta}, \orgaddress{\street{8406 Rue Marie-Anne Gaboury}, \city{Edmonton}, \state{AB} \postcode{T6C 4G9}, \country{Canada}}}

\affil[2]{\orgdiv{Department of Physics}, \orgname{University of Regina}, \orgaddress{\street{3737 Wascana Parkway}, \city{Regina}, \state{SK} \postcode{S4S 0A2}, \country{Canada}}}

\affil[3]{\orgdiv{Department of Physics}, \orgname{University of Alberta}, \orgaddress{\street{11335 Saskatchewan Drive NW}, \city{Edmonton}, \state{AB} \postcode{T6G 2E1}, \country{Canada}}}

\affil[4]{\orgdiv{Instituto de F\'isica Corpuscular}, \orgname{CSIC--Universitat de Val\`encia}, \orgaddress{\street{Catedr\'atico Jos\'e Beltr\'an, 2}, \postcode{46980} \city{Paterna}, \country{Spain}}}

\abstract{Minicharged particles (mCPs), hypothetical free particles with effective electric charges much smaller than the elementary charge, $e$, offer a valuable probe of dark sectors and fundamental physics through several clear experimental signatures.  Various models of physics beyond the Standard Model predict such particles, the existence of which could help elucidate the ongoing mysteries regarding electric charge quantization and the nature of dark matter.  Moreover, a hypothetical scenario involving a small minicharged subcomponent of dark matter has recently been demonstrated as a viable explanation of the anomaly in the 21~cm hydrogen absorption signal reported by the EDGES collaboration.  Although several decades of indirect observations and direct experimental searches for mCPs at particle accelerators have led to severe constraints, a substantial window of the mCP mass--mixing parameter space remains unexplored at the energy frontier accessible to current state-of-the-art accelerators, such as the Large Hadron Collider (LHC).  Consequently, mCPs have remained topical over the years, and new experimental searches at accelerators have been gaining interest.  In this article, we review the theoretical frameworks in which mCPs emerge and their phenomenological implications, the current direct and indirect constraints on mCPs, and the present state of the ongoing and upcoming searches for mCPs at particle accelerators.}


\maketitle

\section{Introduction}\label{sec:intro}
The apparent quantization of electric charge and the nature of dark matter (DM) are among several key unsolved puzzles in modern physics that cannot be explained by the Standard Model (SM) of particle physics in its current formulation.  Many possible explanations for electric charge quantization have been proposed, such as Dirac's demonstration that electric charge quantization is a natural consequence of the existence of a single (yet undiscovered) magnetic monopole~\cite{Dirac1931}.  Alternately, theories of grand unification that embed the hypercharge gauge group into a larger symmetry group have been developed to explain the quantization of the electric charges of the fermions~\cite{Georgi1974,Pati1974}.  Searches for unconfined particles with small and potentially irrational/non-quantized electric charges below the elementary value of electric charge, $e$, provide a stringent test for predictions from these theories and of electric charge quantization in general.  These hypothetical particles, referred to hereafter as minicharged particles (mCPs), are predicted in a variety of beyond the Standard Model (BSM) physics scenarios.

For example, mCPs arise generically within the framework of vector portal dark sector models that extend the SM gauge group to include an additional $U(1)$ gauge symmetry~\cite{Holdom1986_1}.  Dark sector models have been of topical interest in recent years, which unlike traditional particle models of DM that usually propose a single DM particle candidate, instead propose a new dark/hidden sector that could contain its own potentially rich particle content.  Minicharged particles offer a valuable and robust probe of dark sector models, as well as various fundamental physics questions, due to their stability (resulting from electric charge conservation) and relatively well-understood feeble interactions with matter.  Notably, they have a clear, albeit relatively challenging to detect, experimental signature presented as anomalously low ionization or excitation energy deposited in a particle detector.  Minicharged particles have also been invoked in potential resolutions of various anomalies, such as the muon $g-2$ anomaly~\cite{Bai2021} and the anomaly in the $21$~cm hydrogen absorption signal measured by the Experiment to Detect the Global Epoch of Reionization Signature (EDGES) collaboration~\cite{Munoz2018,Berlin2018,Kovetz2018,Liu2019,Aboubrahim2021,Mathur2022,Chu2024}.

Over the last four decades, a substantial amount of experimental effort has been dedicated to the search for mCPs, leading to a plethora of constraints on the corresponding mass--mixing parameter space~\cite{PhysRevLett.48.1649,PhysRevD.35.391,Davidson1991,Akers1995,Prinz1998,Davidson2000,Prinz:2001qz,Gies2006,Chatrchyan2013,Chatrchyan2013_2,Magill2019,Acciarri2020,Ball2020,Plestid2020,Marocco2021,Barak2024,Hayrapetyan2025,Aalbers2025,Alcott2025}.  Additionally, constraints have been derived through the various phenomenological implications that arise from mCPs in precision QED~\cite{Dobroliubov1990,Gluck2007}, stellar astrophysics~\cite{Dobroliubov1990,Mohapatra1990,Davidson1991,Davidson1994,Raffelt1996,Davidson2000}, and cosmology~\cite{Mohapatra1990,Davidson1991,Davidson1994,Davidson2000,Dubovsky2004,MELCHIORRI2007,BEREZHIANI2009,Brust2013,Vogel2014}.  However, despite these abundant experimental and theoretical limitations on the allowed values of the mass and effective electric charge of mCPs, a wide range of masses and charges remain possible.  In particular, a considerable window of the parameter space for $\BigOSI{}{GeV}$ mCP masses remains unexplored at the current energy frontier---a prime target for dedicated search experiments at modern particle accelerator facilities, such as the Large Hadron Collider (LHC).  A host of accelerator experiments dedicated to exploring this unconstrained region of parameter space over the next decade have been proposed in the literature~\cite{Haas2015,Kelly2019,Kim2021,Ball2021,Foroughi2021,Pinfold2791293,Kling2022,Citron2025}.  Three such experiments have been approved to collect data at the LHC's current operational run (Run~3), namely the milliQan and MAPP-1 experiments~\cite{Ball2021,Pinfold2791293} and the FORMOSA demonstrator~\cite{Citron2025}.  

In this mini-review, we cover the current theoretical and experimental landscape of mCPs, with a particular focus on accelerator-based phenomenology and searches. We begin with a review of the theoretical aspects concerning mCPs in Sec.~\ref{sec:mCPthry}. In particular, we discuss the models in which mCPs emerge, their phenomenological implications, and several other important theoretical considerations regarding mCPs. The possibility of mCPs as candidate DM particles is also discussed. In Sec.~\ref{sec:mCPexp}, we provide an overview of the completed, approved and ongoing, and proposed future searches for mCPs at particle accelerators. Additionally, we summarize the projected sensitivities of the main dedicated mCP search experiments at the High-Luminosity LHC (HL-LHC), highlighting their largely complementary coverage and the projected state of the art in accelerator-based mCP searches. Finally, concluding remarks and future perspectives are provided in Sec.~\ref{sec:Conc}. 


\section{Minicharged Particles---Theoretical Considerations}\label{sec:mCPthry}
In this section, we begin with a review of the canonical model that naturally yields mCPs.  Several other models in which unconfined mCPs also emerge are briefly discussed thereafter.  We follow these discussions with an overview of the phenomenological implications of the existence of mCPs and the constraints that arise as a result.  Lastly, we review the status of mCPs as potential DM candidates.

\subsection{Models with Unconfined Minicharged Particles}\label{subsec:mCPmodels}

\subsubsection{Minicharged Particles in Dark Sector Models}\label{subsubsec:mCPinDS}

The standard dark sector model that predicts mCPs was first described in 1986 by Bob Holdom~\cite{Holdom1986_1}.  In this model, a new massless abelian $U(1)$ gauge field, $A'_{\mu}$ (the dark photon\footnote{Also referred to in the literature as a hidden photon or paraphoton.}), kinetically mixes with the SM hypercharge gauge field, $B^{\mu}$. More generally, the additional underlying abelian $U(1)$ gauge symmetry could be a relic from a larger unification gauge group. Additionally, a new massive Dirac fermion ($\chi$) that couples to the dark photon gauge field is included and, hence, is charged under this new $U(1)$ gauge field with an electric charge, $e'$.  The Lagrangian for this model can be written as,
\begin{align}
\mathcal{L} & = \mathcal{L}_{\mathrm{SM}} -\frac{1}{4}  A'_{\mu \nu} A'^{\mu \nu}  + i \bar{\chi} \left( \slashed{\partial} + i e' \slashed{A}' +i m_{\chi} \right) \chi \nonumber \\
& - \frac{\kappa}{2} A'_{\mu \nu} B^{\mu \nu} , \label{EQN:E1}
\end{align}
where $\kappa$ is an arbitrary (potentially irrational) small dimensionless parameter that controls the degree of the kinetic mixing, $m_{\chi}$ is the mass of the dark fermion, and $A'_{\mu \nu}$ is the field strength tensor for the dark photon defined in the usual way as $A'_{\mu \nu} = \partial_{\mu} A'_{\nu} - \partial_{\nu} A'_{\mu} $.  The last term in Eq.~\ref{EQN:E1} containing the kinetic mixing can be eliminated through a field redefinition of the dark photon gauge field, i.e., by diagonalizing the kinetic terms through a shift $A'_{\mu} \Rightarrow A'_{\mu} + \kappa B_{\mu}$.  Applying this field redefinition reveals a coupling between the charged matter field $\chi$ and the SM hypercharge gauge field, apparent in the following Lagrangian,
\begin{align}
\mathcal{L} & = \mathcal{L}_{\mathrm{SM}} -\frac{1}{4}  A'_{\mu \nu} A'^{\mu \nu}  \nonumber \\
& + i \bar{\chi} \left(   \slashed{\partial} + i e' \slashed{A}' - i \kappa e' \slashed{B} + i m_{\chi} \right) \chi. \label{EQN:E2}
\end{align}
Indeed, in the visible sector, the new fermionic field $\chi$ behaves as a field charged under hypercharge with a minicharge of $\kappa e'$, and couplings to the photon and $Z^{0}$ boson of $\kappa e' \cos{\theta_{\mathrm{W}}}$ and $-\kappa e' \sin{\theta_{\mathrm{W}}}$, respectively.  Expressing the effective charge in terms of the elementary charge thus gives $\epsilon \equiv  \kappa e' \cos{\theta_{\mathrm{W}}}/e$.  

\begin{figure}[htb!]
\centering
\includegraphics[width = 8 cm]{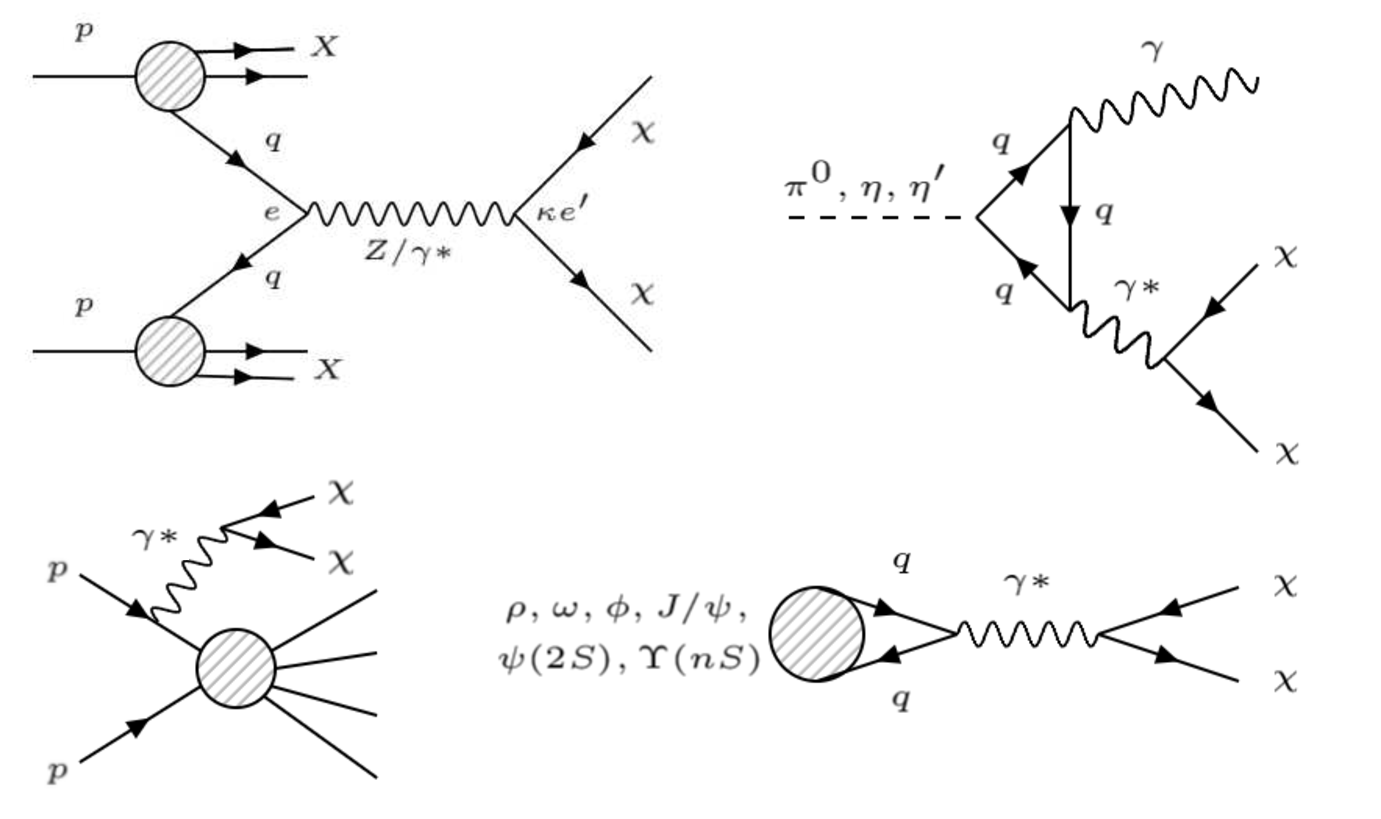}
\caption{Feynman diagrams corresponding to several principal mCP production modes pertinent to their searches at particle accelerators.}\label{Figs:Fig1}
\end{figure}

If the new $U(1)$ gauge symmetry is unbroken, then the matter field that is charged under it is stable.  Furthermore, through the coupling to the SM hypercharge gauge field, a variety of well-understood production mechanisms could lead to abundantly produced mCPs both in the Universe and in the laboratory at particle accelerators.  In the latter scenario, principal pair-production mechanisms include the Drell--Yan process ($q\bar{q}~\rightarrow~\bar{\chi} \chi$); meson decays, such as direct decays of light vector mesons (e.g.,~$\rho~\rightarrow~\bar{\chi} \chi$) and heavy quarkonia (e.g.,~$J/\psi~\rightarrow~\bar{\chi} \chi$), as well as Dalitz decays of pseudoscalar mesons (e.g.,~$\pi^{0}~\rightarrow~\gamma \bar{\chi} \chi$); bremsstrahlung (e.g., $p N~\rightarrow~\bar{\chi} \chi X$); and photoproduction ($\gamma N~\rightarrow~\bar{\chi} \chi X$).  Feynman diagrams corresponding to each of these production mechanisms are provided in Fig.~\ref{Figs:Fig1}.

A generalized version of this model has also been considered, in which the new vector boson that kinetically mixes with the hypercharge gauge field is instead a linear combination of both a massive and a massless state~\cite{Izaguirre2015}. In this ``mixed-phase'' scenario, maximal detector sensitivity is expected in an intermediate $Z'$ mass range, where production remains significant while experimental constraints still permit relatively large kinetic mixing. At low $M_{Z'}$, strong constraints suppress the mixing, whereas at high $M_{Z'}$, the production cross-section decreases. Although this scenario introduces additional constraints with regard to the mass of the dark vector boson, sizeable regions of the parameter space remain unexplored and accessible at contemporary high-energy colliders (i.e., for masses in the range of hundreds of MeVs to hundreds of GeVs). In such a model, an interesting and unique new pair-production mechanism for mCPs emerges through the decay of this new vector boson. This gives rise to a distinct accelerator-unique signature: if the massive dark vector boson is sufficiently boosted (e.g., through jet-associated production), its decay inside or in front of a dedicated downstream detector can result in two mCPs simultaneously traversing the detector~\cite{Izaguirre2015}.  In the event that this unique two-mCP signal is detected, an estimate of the mass of the new vector boson could be obtained.

Another interesting variation of Holdom's kinetic-mixing model has been considered by Brummer et al., in which the dark sector containing the new $U(1)$ gauge symmetry also includes magnetic monopoles~\cite{Brummer2009_2}.  These magnetic monopoles could be, for example, 't Hooft--Polyakov monopoles originating from a spontaneously broken non-Abelian group in the dark sector.  Additionally, their model considers a specific scenario with CP-violating couplings that cause the dark monopoles to acquire an electric charge in the dark sector through the Witten effect, effectively acting as dark dyons~\cite{Brummer2009_2}.  Subsequently, the dark monopoles can be electrically charged under the SM hypercharge gauge group through a natural analog of kinetic mixing (which the authors refer to as ``magnetic mixing'')~\cite{Brummer2009_2}, leading to their appearance as mCPs in the visible sector.  For sufficiently light dark monopoles, searches for their electric minicharges could be performed at current and future experiments~\cite{Brummer2009_2}.

\subsubsection{Minicharged Particles in String Theories and Other Models}\label{subsubsec:mCPinSTandMore}
A handful of other models presented in the literature predict the existence of mCPs.  For example, in certain (super)string theories, such as those with a broken grand unified gauge group, unconfined color singlet mCPs with electric charges of $e/k$ (where $k$ is an integer) emerge~\cite{Wen1985}.  In many string theory models with string scales around $M_{s}\sim10^{11}$~GeV, a lower bound on the effective mCP charge of approximately $10^{-9}\:e$ is predicted~\cite{ABEL2008}.  Moreover, in Ref.~\cite{ABEL2008}, Abel et al. demonstrated how several of the key components of Holdom's model, such as the kinetic-mixing interaction and multiple $U(1)$ gauge groups, can naturally arise in string theoretic extensions of the SM.  Minicharged fermions have also been modelled in self-consistent string theories defined in $5$D with $U(1)$ gauge fields that inhabit a slice of $\mathrm{AdS}_{5}$~\cite{Batell2006}.  Notably, Batell and Gherghetta demonstrated that the massless kinetic-mixing scenario leading to mCPs can be generalized in this particular way of constructing a unified theory in $5$D~\cite{Batell2006}.  Finally, color singlet (under $SU(3)_{C}$) minicharged scalar fields have been discussed in connection with a potential spontaneous $U(1)_{\mathrm{EM}}$ symmetry breaking mechanism~\cite{Ignatiev1979,West2019}.

\subsection{Phenomenological Implications of Minicharged Particles}\label{subsec:mCPpheno}
The existence of scalar or fermionic mCPs would lead to several noteworthy phenomenological implications (several of which have been previously discussed in Ref.~\cite{Dobroliubov1990}, for example).  First, we discuss the effects of mCPs on precision QED measurements.  In the case of light mCPs~($m_{\chi}~\lesssim~100$~MeV), a significant contribution to the photon vacuum polarization would arise due to their creation in vacuum bubbles.  Consequently, there would be an alteration in the Lamb shift of atoms at the one-loop level.  Thus, precision measurements of the Lamb shift of hydrogen atoms can be used to constrain the mass--mixing parameter space of light mCPs.  In this fashion, mass-independent upper bounds on the effective charge of light fermionic mCPs ($m_{\chi} \lesssim 1$~keV) of~$Q_{\chi}~\lesssim~10^{-4}\:e$ were established~\cite{Gluck2007}.  In the case of scalar mCPs, a two-fold improvement on these bounds was obtained~\cite{Gluck2007}.  Similarly, mCPs would affect the anomalous magnetic moments of the electron and muon, $g_{e} - 2$ and $g_{\mu} - 2$, respectively, at the two-loop level.  Indeed, extremely precise measurements of $g_{e} - 2$ and $g_{\mu} - 2$ have been used to constrain mCPs~\cite{Dobroliubov1990}; however, these bounds are dominated by those obtained from measurements of the Lamb shift due to the increased interaction order.

Second, there are several important implications for astrophysics concerning stellar evolution in particular.  Since sufficiently light mCPs would be produced in the hot dense plasma interior of stars and could escape for $\epsilon \lesssim 10^{-8}$~\cite{Davidson2000}, stellar energy-loss arguments can be applied to constrain mCPs~\cite{Dobroliubov1990,Mohapatra1990,Davidson1991,Davidson1994,Raffelt1996,Davidson2000}.  In the case of horizontal branch stars, this additional energy-loss mechanism due to the emission of mCPs produced via plasmon decays would result in enhanced helium burning and a shortened stellar lifetime that would affect observations of the population density of horizontal branch stars in globular clusters~\cite{Davidson1991,Davidson1994,Raffelt1996,Davidson2000}.  The same argument can be applied to white dwarfs, which would cool faster than normal due to the emission of mCPs~\cite{Dobroliubov1990,Davidson1991,Davidson2000}; the known average age of a white dwarf thus constrains the possible effective charge of mCPs.  In both cases, strong upper bounds on the effective charge of $Q_{\chi} \sim 10^{-13}\: e$ were obtained for very light mCPs ($m_{\chi} \lesssim 10$~keV)~\cite{Dobroliubov1990,Davidson2000}.  Supernovae are also susceptible to the energy-loss argument, which has been applied to observations of the neutrino pulse duration from Supernova 1987A in several studies~\cite{Mohapatra1990,Davidson1994,Davidson2000}.  The region of the mass--mixing parameter space from~$10^{-9}\lesssim~\epsilon~\lesssim~10^{-7}$ for mCP masses of $m_{\chi} \lesssim 10$~MeV has been excluded as a result~\cite{Mohapatra1990,Davidson2000}.  For mCPs with effective charges above $ \sim10^{-7}\:e$, the interaction strength is large enough that they would not escape red giant stars and would instead facilitate heat transport~\cite{Dobroliubov1990}.  An argument based on such considerations has been applied in Ref.~\cite{Dobroliubov1990}, resulting in lower bounds of $\epsilon \gtrsim 2 \times 10^{-3}$ and $\epsilon~\gtrsim~4~\times~10^{-3}\exp{\left(-m_{\chi}/20\:\text{keV}\right)}$, for mCP masses of $m_{\chi}~<~10$~keV and $m_{\chi}~>~10$~keV, respectively.

Lastly, the potential presence of mCPs in the early Universe has several implications for cosmology, regarding Big Bang nucleosynthesis (BBN)~\cite{Mohapatra1990,Davidson1991,Davidson1994,Davidson2000,BEREZHIANI2009,Vogel2014} and the cosmic microwave background (CMB) anisotropies~\cite{Dubovsky2004,MELCHIORRI2007,Brust2013,Vogel2014,Adshead_2022}.  The production of dark photons and mCPs in the early Universe would act as a form of dark radiation that contributes to the overall radiation energy density, which is typically parameterized by the effective number of different neutrino species, $N_{\mathrm{eff}}$.  Light elements produced in the early Universe, such as $^{4}\mathrm{He}$ and $^{2}\mathrm{D}$, have a strong dependence on $N_{\mathrm{eff}}$ and, hence, on the overall radiation energy density.  The SM prediction for the effective number of neutrino species is $N_{\mathrm{eff}} \simeq 3.046$~\cite{MANGANO2002,MANGANO2005}.  Based on astrophysical measurements of the relic abundance of primordial $^{4}\mathrm{He}$ performed by the Wilkinson Microwave Anisotropy Probe (WMAP), for example, an estimate on the effective number of neutrino species of $N_{\mathrm{eff}} = 3.14^{+0.70}_{-0.65}$ was obtained at the $68$\% confidence level (CL)~\cite{CYBURT2005}.  Deviations in the theoretically predicted value of $N_{\mathrm{eff}}$ occur if light mCPs existed in the early Universe, thereby allowing one to use precision measurements of $N_{\mathrm{eff}}$ to indirectly constrain the allowed values of the mCP mass and effective charge.  Similarly, the existence of mCPs in the early Universe would leave an imprint on the CMB anisotropies; thus, precise measurements of the CMB anisotropy spectrum (e.g., the full-mission data from Planck~\cite{Planck2018cosmo}) can also be used to obtain indirect limits on mCPs.  The most recent $2\sigma$ upper bound established from this dataset limits sub-GeV mCPs to effective charges below $\sim10^{-9}$--$10^{-7}\:e$, depending on the mCP mass~\cite{Adshead_2022}. For this mass range, these results slightly improve the previous indirect limits~\cite{Vogel2014} determined from the Planck 2013 data release~\cite{Planck2013cosmo}; however, for $\BigOSI{}{GeV}$ mCP masses, a sizeable increase in the upper mass limit was obtained, extending the previous constraints beyond $1$~GeV up to $\sim5$~GeV. Complementary indirect bounds on $10$--$150$~MeV mCPs determined from measurements of the CMB have also been reported by Brust et al.~\cite{Brust2013}.

\subsection{Minicharged Particles as Dark Matter Candidates}\label{subsec:mCPasDM}
A multitude of attempts to explain dark matter through extensions of the SM have been developed since its inception, resulting in numerous hypothetical DM particle candidates.  The specific criteria for such a particle must be consistent with our current knowledge of the properties of DM; hence, particle candidates of DM must be massive, stable/long-lived over cosmological timescales~\cite{Audren_2014}, and very weakly interacting with normal matter~\cite{PhysRevLett.115.071304}.  As previously mentioned, if the mCPs are charged in the dark sector under an unbroken $U(1)$ gauge group, then they are stable; in this framework, Daido et al.~\cite{Daido2017} demonstrated that gauge coupling unification is improved, with minicharged DM emerging as a natural consequence of unification.  Moreover, such mCPs satisfy the aforementioned properties to be considered suitable candidates for non-baryonic DM, although there are strong constraints on the viable values of effective electric charge~\cite{PhysRevD.83.063509,PandaX2023}.

Beyond these basic criteria for a particle model of DM, another key requirement for any DM model is some mechanism(s) for generating the observed DM abundance in the Universe.  Several studies have demonstrated the viability of both thermal~\cite{Hall2010,Chu2012,Dvorkin2019} and non-thermal~\cite{Bogorad2021} mechanisms for generating minicharged DM (mC-DM) in the early Universe.  Furthermore, mC-DM models must obey the current direct and indirect constraints on mCPs.  In particular, the previously discussed constraints derived from CMB data place a severe restriction on the allowed relic density of mC-DM~\cite{Dubovsky2004,Dolgov2013,Kovetz2018,Boddy2018,Aboubrahim2021}.  Based on the CMB data released from Planck in 2015~\cite{Planck2015cosmo,Planck2015cosmoII}, mC-DM is restricted to a scenario in which mCPs can only comprise a small subcomponent of the total DM relic density ($f_{\chi}\lesssim 0.4\%$)~\cite{Kovetz2018,Boddy2018,Aboubrahim2021}.  In Ref.~\cite{Aboubrahim2021}, the authors compute the relic density of mC-DM and demonstrate that an amount that is consistent with these constraints from CMB data can be obtained for an experimentally allowed range of mCP parameters.

A scenario in which the mC-DM subcomponent interacts ``strongly''\footnote{Here, `strongly' refers to dark matter particles that have larger-than-weak cross-sections with Standard Model particles rather than interactions mediated by the strong nuclear force (QCD).} with SM particles, so-called minicharged strongly interacting DM (mC-SIDM), has also been discussed throughout the literature~\cite{Emken2019}.  Numerous direct detection experiments have performed searches for SIDM by looking for evidence of DM--nucleus interactions in the detector media, e.g., elastic DM--electron scattering events.  However, the sensitivity of these searches is limited to a band of parameter space, above which the SIDM interacts too frequently prior to reaching the detector such that it cannot leave a measurable signal in the detector and below which the interactions are too weak to produce a sufficient number of signal events.  The upper bound is determined by the ``critical cross-section''~\cite{Emken2019}, which cannot be avoided by ground-based direct detection experiments.  Consequently, searches for SIDM have also been performed at high altitudes, enabling the possibility of probing interaction strengths in the region above the critical cross-section.  In addition to all the previously discussed constraints on mCPs, the constraints from ground-based direct DM detection experiments~\cite{Emken2019}, rocket-based experiments~\cite{Erickcek2007}, and nuclear recoil experiments performed at high altitudes~\cite{Rich1987} further restrict the possibility of mC-SIDM. Complementary constraints may also be established by space-based detectors such as the Alpha Magnetic Spectrometer (AMS), which could examine its existing cosmic-ray data for possible signatures of mC-SIDM in regions of parameter space inaccessible to ground- or high-altitude experiments~\cite{AGUILAR20211}. However, for the maximal viable mC-DM subcomponent of $f_{\chi} = 0.4\%$, a unique window of the mC-SIDM parameter space remains unconstrained and accessible to contemporary accelerator experiments (see Figure~11 of Ref.~\cite{Emken2019}).  Additionally, it is noteworthy to mention that accelerator-based searches have a distinct advantage, as their corresponding constraints on mC-SIDM do not depend on $f_{\chi}$ and are not limited by the critical cross-section.

Lastly, several studies~\cite{Munoz2018,Berlin2018,Kovetz2018,Liu2019,Aboubrahim2021,Mathur2022,Chu2024} have demonstrated that a small minicharged subcomponent of DM could resolve the anomaly in the $21$~cm hydrogen absorption spectrum (centered at a redshift of $z \sim 17$ and covering $z \sim 15$--$20$) reported by the EDGES collaboration in 2018~\cite{Bowman2018}.  Specifically, the anomaly indicates a significant increase in absorption---at least double that of the largest predictions~\cite{Bowman2018}.  To explain this feature, either cooling of the $\mathrm{H}$ gas or radiative heating of the CMB is necessary.  Thus, if it is indeed a real result (and not the consequence of an incomplete study of the systematics, for example), then new physics appears to be required to explain this anomaly.  Through the introduction of a small minicharged component of DM that (feebly) interacts with baryons through Coulomb interactions, one can obtain the cooling of hydrogen required to resolve the absorption anomaly.  Furthermore, two independent analyses have shown that even at the small upper limit of $f_{\chi} = 0.4\%$ for the mC-DM fraction, sufficient cooling to resolve the EDGES anomaly can be obtained for a substantial window of unconstrained mCP parameter space~\cite{Berlin2018,Kovetz2018}.  In the more recent of these studies, this corresponds to a maximal range of viable mC-DM particle masses and effective charges from approximately $0.5$ to $35$~MeV and $6\times10^{-6}$ to $2\times10^{-4}\:e$ (depending on the mass), respectively~\cite{Kovetz2018}; a region that could be probed by future searches for mCPs, including those performed at current accelerator facilities. Despite this region being technically ruled out by indirect $2\sigma$ upper bounds arising from precise CMB measurements of $N_{\mathrm{eff}}$, these bounds are model-dependent, and the viable parameter space can be extended to larger mC-DM particle masses above this bound (up to a few hundred GeV) by including a long-range dark force between the primary cold dark matter component and the mC-DM subcomponent~\cite{Liu2019}. At the maximal fraction $f_{\chi} = 0.4\%$, mC-DM sufficiently coupled to baryons at recombination could also settle the $2\sigma$ BBN--CMB tension in the baryon energy density~\cite{Kovetz2018}.


We note that this mCP-based resolution of the EDGES anomaly has been criticized, e.g., in Ref.~\cite{Sarbinowski2019}, which demonstrated that, under the assumption of thermal freeze-out, this scenario is excluded by cosmological bounds on relativistic degrees of freedom. However, subsequent studies have shown that this critique can be circumvented in non-minimal models: one approach mentioned above introduces a long-range interaction between the mC-DM subcomponent and the cold dark matter, extending the allowed parameter space and loosening cosmological constraints~\cite{Liu2019}, while another allows freeze-out to include annihilation into neutrinos, thereby evading the concerns that ruled out the simplest scenario~\cite{Chu2024}. These extensions indicate that mC-DM remains a viable candidate for explaining the EDGES signal within more complex dark sector frameworks, albeit with some fine-tuning required.

\section{Minicharged Particle Searches at Accelerator Experiments}\label{sec:mCPexp}
A plethora of stringent searches for mCPs have already been performed at various particle accelerator experiments, and many contemporary searches have been proposed in the last decade (several of which have been approved).  This section begins with a review of past searches for mCPs at particle accelerators, presented in chronological order, followed by discussions of approved and ongoing searches, and proposed future searches at accelerators.

\subsection{Past Searches at Accelerators}\label{subsec:pastsrch}
The first search for the exclusive production of unconfined and stable fractionally charged particles with charges as low as~$Q_{\chi}~=~e/3$, reported in 1982, was the Free Quark Search (FQS)~\cite{PhysRevLett.48.1649} at the Stanford Linear Accelerator Center (SLAC)\footnote{Now called the SLAC National Accelerator Laboratory.} Positron--Electron Project (PEP).  The PEP accelerator was an electron--positron collider that reached center-of-mass energies of $29$~GeV~\cite{4329104}.  The $e^{+}e^{-}$ collision products were studied by the FQS experiment to search for fractionally charged particles produced in exclusive two-body final states~\cite{PhysRevLett.48.1649}.  No such states were detected~\cite{PhysRevLett.48.1649}.  Consequently, the first exclusion bounds for leptons with fractional charges as low as~$Q_{\chi}~=~e/3$ were reported, reaching masses as large as approximately $14$~GeV (at the $90$\% CL)~\cite{PhysRevLett.48.1649}.

Several years later, in 1987, Golowich and Robinett considered the photoproduction of mCPs at past beam-dump experiments (such as the Fermilab E613 neutrino beam-dump experiment~\cite{Ball:1980ojt}) through the process $\gamma + N \rightarrow \chi + \bar{\chi} + X$, where $N$ represents a nucleon~\cite{PhysRevD.35.391}.  Resulting limits on mCPs with masses of $1~\lesssim~m_{\chi}~\lesssim~200$~MeV and effective charges reaching down to~$\sim2\times10^{-2}\:e$ were established~\cite{PhysRevD.35.391}.  Subsequently, a comprehensive study of experimental limits on mCPs performed by Davidson et al. was presented in 1991~\cite{Davidson1991}.  A wide range of limits arising from various sources, including direct accelerator searches, stellar astrophysics, and cosmology, were considered~\cite{Davidson1991}.  Regarding the accelerator searches in particular, the authors reinterpreted the data from several experiments: the Large Electron--Positron (LEP) Collider~\cite{osti_6078475}, the SLAC Beam Dump experiment~\cite{Rothenberg1972aSF}, and the Anomalous Single Photon (ASP) experiment at SLAC~\cite{PhysRevLett.58.1711}.  In combination with the other sources of limits considered in their study, significant constraints on the mCP mass--mixing parameter space were placed~\cite{Davidson1991}.

A dedicated search for mCPs at SLAC---the Millicharged Particle Search (mQ) experiment~\cite{Prinz1998}---was proposed and approved in 1993.  This high-energy electron beam-dump experiment ran from 1994 to 1998, using a scintillation detector dedicated to searching for the anomalously low-energy ionization signature of mCPs.  While the SLAC mQ experiment was collecting data, the Omni-Purpose Apparatus for LEP (OPAL) collaboration published the results of their search for fractionally charged particles using data collected from 1991 to 1993 at the LEP collider (totaling an integrated luminosity of $74$~pb$^{-1}$)~\cite{Akers1995}.  The OPAL search excluded mCPs with charges $Q_{\chi} \geq 2e/3$ for masses as large as $84$~GeV~\cite{Akers1995}.  The final results of the SLAC mQ experiment, reported several years later, detected no clear experimental signatures consistent with mCPs and excluded minicharges as low as $\sim10^{-5}\:e$ (depending on the mCP mass) over a range of mCP masses from approximately $10^{-8}$~MeV to $100$~MeV~\cite{Prinz1998,Prinz:2001qz}.  A summary of the past mCP searches at accelerators discussed so far, including numerous additional sources of constraints on mCPs and various improvements to several different limits, can be found in Ref.~\cite{Davidson2000} (see Figure~1 for a corresponding plot of the excluded mass--mixing parameter space).

Throughout the 2000s, efforts toward the search for mCPs at accelerators slowed down significantly.  As far as we are aware, there was only one additional accelerator-based limit published in this time, which employed a novel method to exploit the intense electric fields found in accelerator cavities to constrain mCPs produced via the Schwinger mechanism~\cite{Gies2006}.  If produced, the mCPs would contribute to the overall energy loss of the cavity, which, for a certain range of values of the mCP mass and mixing parameter, could result in a measurable effect on the quality factor of the cavity.  In particular, the superconducting cavity of the Tera Electronvolt Superconducting Linear Accelerator (TESLA), with an accelerating gradient of $\varepsilon_0=25$~MV/m and a quality factor of $Q=10^{10}$~\cite{LILJE2004}, was considered.  No significant deviation in the $Q$ factor was ever reported; consequently, exclusion bounds on ultralight mCPs with masses of $10^{-6} \leq m_{\chi} \lesssim  5\times10^{-2}$~eV and charges as low as $\sim 9 \times 10^{-7} \:e$ could be placed~\cite{Gies2006}.

The next set of searches for unconfined fractionally charged particles was performed by the Compact Muon Solenoid (CMS) experiment at the LHC.  The results of two such searches based on data collected at the LHC's Run~1 were reported in 2013~\cite{Chatrchyan2013,Chatrchyan2013_2}. The first study analyzed a total of $5.0$~fb$^{-1}$ worth of data collected in $pp$ collisions at a center-of-mass energy of $\sqrt{s} = 7$~TeV~\cite{Chatrchyan2013}.  In the second study, the full Run~1 dataset collected by the CMS experiment was analyzed, i.e., data collected from $\sqrt{s}=8$~TeV $pp$ collisions during the second half of Run~1 were included in the analysis, amounting to an additional $18.8$~fb$^{-1}$ of integrated luminosity~\cite{Chatrchyan2013_2}.  The final results of these two analyses exclude, at the $95$\% CL, mCPs with masses of $100~\lesssim~m_{\chi}~<~480$~GeV for electric charges of $Q_{\chi}~=~\pm 2e/3$~\cite{Chatrchyan2013,Chatrchyan2013_2}.

Finally, a handful of mCP searches at accelerators have been published in the last seven years.  In 2019, new limits on mCPs were reported based on reinterpretations of the electron-scattering data released by two past neutrino experiments~\cite{Magill2019}: the Liquid Scintillator Neutrino Detector (LSND) and the Mini Booster Neutrino Experiment (MiniBooNE).  Specifically, the sensitivities of these experiments to mCPs were established by studying low-energy electron recoils as a probe for mCPs based on elastic $e^{-}$--$\chi$ scattering.  Notably, the corresponding analysis of the LSND dataset~\cite{Auerbach2001}---collected using a $0.798$~GeV proton beam dump and a total of $1.7 \times 10^{23}$ protons on target (POT)---established new sensitivity (at the $95$\% CL) below the bounds set by the SLAC mQ experiment for light mCPs ($5~\lesssim~m_{\chi}~\lesssim~35$~MeV)~\cite{Magill2019}.  Furthermore, in their analyses of the MiniBooNE data, which considered two datasets collected using Fermilab's $8.9$~GeV proton beam (neutrino and antineutrino runs from a total sample of $2.41~\times~10^{21}$~POT~\cite{Aguilar2018} and electron-recoil data collected from $1.86 \times 10^{20}$~POT~\cite{Aguilar2018_2}), new constraints on mCPs were established for masses of $100 \lesssim m_{\chi} \lesssim 180$~MeV~\cite{Magill2019}.

The following year, the Argon Neutrino Teststand (ArgoNeuT) experiment~\cite{Anderson2012} reported the results of a similar search for mCPs based on the detection of electron recoils~\cite{Acciarri2020}.  This experiment exposed a $175$~L liquid argon time projection chamber (LArTPC) to Fermilab's Neutrinos at the Main Injector (NuMI) beam~\cite{ADAMSON2016} from September 2009 to February 2010~\cite{Acciarri2020}.  Despite the short exposure time of the ArgoNeuT detector (amounting to $1.0 \times 10^{20}$~POT), world-leading constraints were placed at the $95$\% CL on mCPs with masses ranging from $0.1 \lesssim m_{\chi} \lesssim 5$~GeV and effective charges from $\sim4\times10^{-3}\:e$ to $2\times10^{-1}\:e$~\cite{Acciarri2020}.  

Several months later, the first dedicated search for mCPs with charges below $0.1\:e$ performed at a hadron collider was reported by the milliQan collaboration~\cite{Ball2020}.  A total of $37.5$~fb$^{-1}$ of integrated luminosity from $pp$ collisions at $\sqrt{s}=13$~TeV was collected by their scintillator-based demonstrator detector deployed in 2018 during the LHC's Run~2~\cite{Ball2020}.  The detector was located $33$~m from the CMS interaction point (IP5) and at an azimuthal angle of $\phi \sim 43^{\circ}$ in the CMS coordinate system.  Their analysis considered pair-production of mCPs via the Drell--Yan mechanism and a variety of meson decays (e.g., $\rho \rightarrow \chi \bar{\chi}$ and $\pi^{0} \rightarrow \gamma \chi \bar{\chi}$).  No clear signal events consistent with mCPs were observed.  Consequent bounds on mCPs were set at the $95$\% CL, resulting in new sensitivity over the ArgoNeuT limits for mCPs with masses ranging from $0.7 \lesssim m_{\chi} \lesssim 2$~GeV and $2.5 \lesssim m_{\chi} \lesssim 4$~GeV~\cite{Ball2020}.

Bounds on mCPs derived from past data collected in 1982 by the Big European Bubble Chamber (BEBC) located at the CERN WA66 beam-dump experiment~\cite{1986253} were published in 2021~\cite{Marocco2021}.  In this analysis, Marocco and Sarkar analyzed the corresponding electron-scattering data for an excess of recoils due to scattering with mCPs, which could potentially be produced copiously via the decays of mesons created in the $400$~GeV proton beam dump~\cite{Marocco2021}.  No excess was reported; thus, limits on mCPs were established (at the $90$\% CL), which excluded masses and charges ranging from $10^{-3} \lesssim m_{\chi} \lesssim 4$~GeV and $3 \times10^{-4} \: e \lesssim Q_{\chi} \lesssim 0.3 \: e$, respectively~\cite{Marocco2021}.  Additionally, the authors derived similar limits based on an analysis of the data from the CHARM~II experiment~\cite{DEWINTER1989}.  However, for all values of the mCP mass covered in their study, these limits were surpassed by those set using the data from the BEBC detector~\cite{Marocco2021}.

In 2024, the SENSEI collaboration reported results from a search for mCPs produced in the NuMI beam at Fermilab~\cite{Barak2024}. The analysis utilized data collected in 2020 with silicon Skipper CCD detectors installed in the MINOS cavern, downstream of the NuMI graphite target. In this beam-dump configuration, ultrarelativistic mCPs could be generated in proton--target collisions and subsequently traverse shielding material to reach the detector. The search targeted ionization signals corresponding to $3$--$6$ electrons in single or adjacent pixels, accounting for charge diffusion, as expected from the passage of a minimally ionizing particle with a small electric charge. No candidate events were observed above the expected background. Consequently, new $95\%$ CL exclusion limits were derived, achieving world-leading sensitivity for mCP masses in the range $30 \lesssim m_{\chi} \lesssim 380$~MeV~\cite{Barak2024}. Notably, this study established the feasibility of combining low-threshold silicon detectors with high-intensity proton beam facilities to probe feebly ionizing states.

Following the SENSEI results, the CMS collaboration extended their Run-1 search for fractionally charged particles using LHC Run-2 data~\cite{Hayrapetyan2025}. The analysis used $138~\mathrm{fb}^{-1}$ of proton--proton collision data collected at a center-of-mass energy of $\sqrt{s} = 13$~TeV between 2016 and 2018. This study targeted particles with charges between $e/3$ and $0.9 \:e$, extending previous coverage for $Q_{\chi}= 2e/3$. No significant excess above expected backgrounds was observed, and $95\%$ CL limits were derived, excluding mCPs with masses up to $640$~GeV and charges down to $e/3$, representing the most stringent constraints to date for mCPs in this range of effective charges.

Lastly, the milliQan collaboration recently reported results from proton--proton collisions at $\sqrt{s}=13.6$~TeV using the full milliQan Run~3 bar detector~\cite{Alcott2025}. The analysis utilized $124.7 \pm 3.8$~fb$^{-1}$ of data collected in 2023--24, corresponding to a portion of the total projected LHC Run~3 luminosity. This constitutes the first dedicated search for mCPs at this energy. The observed event rates were consistent with background expectations; consequently, the collaboration established $95\%$~CL exclusion limits on mCPs, with new sensitivity achieved for masses and effective charges $m_\chi \ge 0.45$~GeV and $Q_{\chi} \leq 0.24 \: e$, respectively~\cite{Alcott2025}.

\begin{figure}[htb!]
\centering
\includegraphics[width = 7.6 cm]{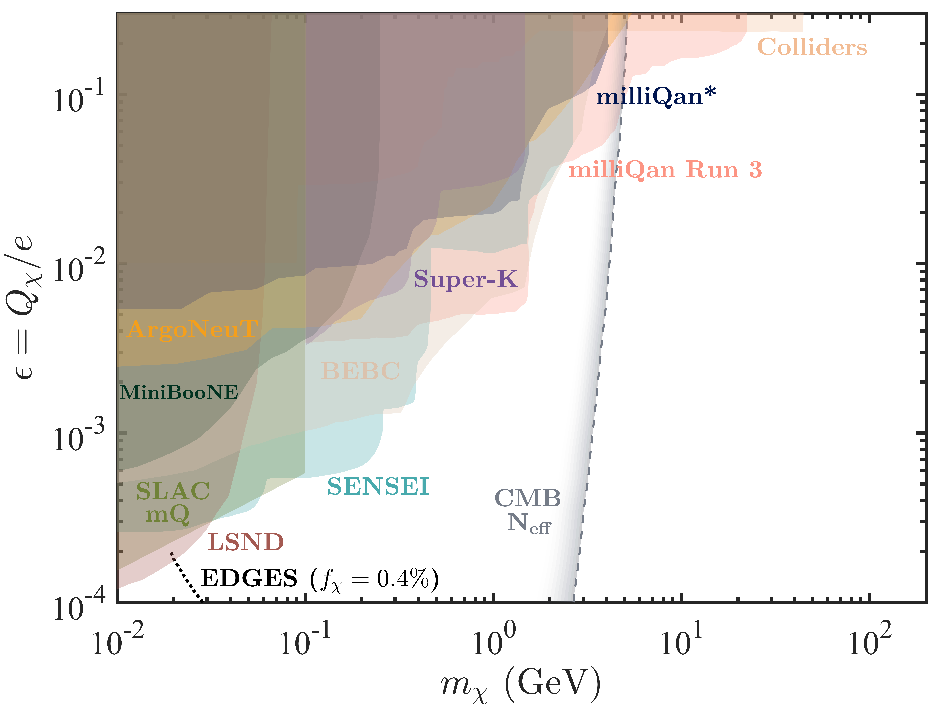}
\caption{The region of the mCP mass--mixing parameter space primarily of interest to particle accelerator experiments at present, with the associated exclusion bounds currently available shown as the shaded regions~\cite{Prinz1998,Davidson2000,Magill2019,Acciarri2020,Ball2020,Plestid2020,Marocco2021,Barak2024,Alcott2025}.  Limits from CHARM~II~\cite{Marocco2021} and LUX-ZEPLIN~\cite{Aalbers2025} are omitted, as discussed in the main text.  All these excluded regions correspond to limits established at the $95\%$ CL, except for the BEBC bounds~\cite{Marocco2021}, which were set at the $90\%$ CL.  The most stringent indirect $2\sigma$ upper limits determined from measurements of the effective number of different neutrino species ($N_{\mathrm{eff}}$) from the CMB~\cite{Adshead_2022} and the maximally allowed region associated with a potential resolution of the EDGES anomaly~\cite{Kovetz2018} are represented with dashed and dotted lines, respectively.} \label{Figs:Fig2}
\end{figure}

Notwithstanding all of the constraints on the mCP mass--mixing parameter space discussed here, a significant region accessible to modern accelerator experiments remains unexplored (corresponding to a range of mCP masses from approximately $0.1$ to $100$~GeV).  A comprehensive exclusion plot of this region of the parameter space is provided in Fig.~\ref{Figs:Fig2}.  The limits on mCPs derived from the CHARM~II dataset reported in the Marocco and Sarkar study~\cite{Marocco2021} are not displayed since, as mentioned, they are surpassed by the additional limits derived in their study using the BEBC data. We also note that $90\%$~CL limits from LUX-ZEPLIN~\cite{Aalbers2025}, derived from a search for atmospheric mCPs produced in cosmic-ray interactions, are not shown in the figure, as they are largely overtaken by the more stringent accelerator-based constraints from SENSEI and milliQan Run~3. The maximally allowed region ($f_{\chi}=0.4$\%) associated with an mCP-based resolution of the EDGES anomaly~\cite{Kovetz2018}, indirect $2\sigma$ upper bounds arising from precise measurements of the effective number of different neutrino species from the CMB~\cite{Adshead_2022}, and additional limits obtained from analyzing the data collected by Super-Kamiokande (Super-K) for mCPs produced by cosmic rays colliding with the atmosphere~\cite{Plestid2020} are also shown.  Several dedicated experiments that aim to explore this unexcluded window of the parameter space have recently been approved, which we now turn to.

\subsection{Approved and Ongoing Searches at Accelerators}\label{subsec:ApprvSrch}
Numerous experiments able to perform searches for mCPs have been approved to run at various accelerators currently in operation.

\subsubsection{The SBN Program}
At Fermilab, the international Short-Baseline Neutrino (SBN) Program~\cite{Machado2019}---proposed in 2015~\cite{Acciarri_2015} and granted Stage~1 approval by the US Department of Energy in the same year---can provide several new searches for mCPs.  Three LArTPC neutrino detectors, each placed at different distances $d$ from the source, comprise the SBN Program: the Short-Baseline Near Detector (SBND), a $112$-ton detector located the closest to the neutrino source ($d~=~110$~m)~\cite{McConkey_2017}; the MicroBooNE detector, a $170$-ton detector located at an intermediate distance from the source ($d~=~470$~m)~\cite{Acciarri_2017}; and the ICARUS-T600 detector, a $760$-ton detector located the furthest from the source~($d~=~600$~m)~\cite{Rubbia_2011}.  All three experiments have the capacity to provide competitive sensitivities to mCPs that surpass the present limits (for mCP masses in the order of hundreds of MeVs).  In fact, the MicroBooNE experiment~\cite{Acciarri_2017} has already completed its run (from 2015 to 2021) and could provide new constraints on mCPs in the near future.  Projected limits on the mass and effective charge of mCPs based on the MicroBooNE liquid argon detector (assuming $1.32 \times 10^{21}$~POT) have been derived in Ref.~\cite{Magill2019}.  Notably, new constraints (set at the $95$\% CL) are predicted for mCPs with masses in the range from $100 \lesssim m_{\chi} \lesssim 250$~MeV~\cite{Magill2019}.  Additionally, the authors derived projected bounds on mCPs based on the SBND, which has been fully installed and commissioned, and has been taking neutrino beam data since summer 2024~\cite{AlvarezGarrote:2024gk}.  Considering a total data volume of $6.6 \times 10^{20}$~POT collected at the SBND, Magill et al. reported marginal improvements over their estimated bounds for the MicroBooNE experiment, reaching effective charges as low as $\sim10^{-4}\:e$~\cite{Magill2019}. While the SBN Program is driven mainly by the low-energy Booster Neutrino Beam (BNB), ICARUS's location also exposes it to a considerable flux of particles from the $120$~GeV high-energy NuMI beamline, and ICARUS has already recorded events from both beams, providing potential sensitivity to higher-mass mCPs with existing or future datasets. Despite this, no searches or projections for mCPs have yet been reported for the ICARUS-T600 detector.

\subsubsection{The $\mathrm{NA}64e$ Experiment}
The $\mathrm{NA}64e$ experiment~\cite{Crivelli2023} was proposed in 2014 and approved by CERN in March 2016. It is a fixed-target missing-energy experiment operating at the CERN Super Proton Synchrotron (SPS), using a high-intensity $100$~GeV electron beam impinging on an active electromagnetic calorimeter target. The experiment is primarily designed to search for sub-GeV dark sector particles produced via electron--nucleus scattering. The $\mathrm{NA}64e$ experiment has sensitivity to mCPs produced in the electron beam dump through decays of vector mesons and bremsstrahlung-like events. According to recent collaboration summaries presented in 2025, the experiment has collected a total of $\sim2.5 \times 10^{12}$ electrons on target (EOT) from data-taking runs spanning the years 2016--2025, with further data taking ongoing. Although no dedicated analysis of mCPs in the collected dataset has yet been reported, projected sensitivities have been reported in a recent study~\cite{Arefyeva2022}. Utilizing a projected $5 \times 10^{12}$ EOT, $\mathrm{NA}64e$ is expected to achieve a $95\%$ CL sensitivity that probes previously unexplored regions of the mCP parameter space, spanning a wide range of couplings and reaching as low as $\epsilon \sim 5 \times 10^{-6}$ for sub-MeV mCPs~\cite{Arefyeva2022}.

\subsubsection{The LDMX Experiment}
The Light Dark Matter eXperiment (LDMX)~\cite{Akesson2018}, proposed in 2018, received Stage~1 approval in 2020.  LDMX is an ultra-short baseline beam-dump experiment that targets new particles with masses in the sub-GeV range that interact with electrons by exploiting a missing momentum/energy technique~\cite{Akesson2018}.  The proposed detector comprises two tracking systems (for electron tagging and recoils), a thin tungsten target placed between the trackers, and an electromagnetic calorimeter surrounded by a large hadronic calorimeter~\cite{Akesson2018}.  Several multi-GeV electron beam facilities were initially considered as potential locations to host LDMX~\cite{Akesson2018}, such as the DArk Sector Experiments at LCLS-II (DASEL) facility at SLAC ($4$--$8$~GeV)~\cite{Raubenheimer2018}, the Continuous Electron Beam Accelerator Facility (CEBAF) at Jefferson Lab ($11$~GeV)~\cite{Leeman2001}, or the proposed electron beam facility at CERN ($16$~GeV)~\cite{Akesson2018primary}. Recent plans now specify that LDMX will be installed in End Station A at SLAC, employing an $8$~GeV electron beam derived from LCLS-II~\cite{Akesson2025}. Three running phases of LDMX have been defined: (1) an early-running phase using an ECal as target (EaT) analysis ($3 \times 10^{13}$~EOT); (2) a pilot run ($4 \times 10^{14}$~EOT); and (3) the full planned exposure targeting $\sim 10^{16} $~EOT, providing the complete LDMX sensitivity~\cite{Akesson2025}.  Projected bounds for LDMX on the mCP mass--mixing parameter space have been published in Ref.~\cite{Berlin2019}.  Notably, LDMX has exceptional sensitivity to sub-GeV mCPs---able to probe a substantial portion of the region associated with a resolution of the EDGES anomaly for $f_{\chi}=0.4\%$.


\subsubsection{The SUBMET Experiment}
The SUB-Millicharge ExperimenT (SUBMET), a proton fixed-target experiment aimed at mCP searches using the $30$~GeV proton beam accelerated in the main ring at the Japan Proton Accelerator Research Complex (J-PARC)~\cite{Choi2020}, received Stage~2 approval in June 2023. A MAPP/milliQan-like detector comprising multiple layers of stacked scintillator bars, each individually coupled to a photomultiplier tube (PMT), was installed in 2024 and has since been taking data to search for mCPs produced principally via the decays of neutral mesons~\cite{Kim2021}; based on $30$~GeV proton--fixed-target collisions, mCPs produced from the decays of primary neutral mesons as heavy as the $\psi\left(2S\right)$ can be studied. Considering a double-layer detector design located $30$~m underground and $280$~m from the target, and a total sample of $N_{\mathrm{POT}}~=~10^{22}$ collected over three years, a projected sensitivity that surpasses the bounds set by both the BEBC and SLAC mQ experiments (over the mass range covered) was reported~\cite{Kim2021}. Notably, excellent sensitivity projections were obtained for light mCPs with masses below $\sim0.2$~GeV, reaching effective charges as low as $\sim 5 \times 10^{-5}\:e$ (at the $95\%$ CL).

\subsubsection{The milliQan Experiment}
At the LHC, three dedicated searches for mCPs have been approved by the CERN Research Board for data taking during Run~3: the milliQan experiment, the FORMOSA demonstrator, and the MoEDAL-MAPP (MoEDAL's Apparatus for Penetrating Particles) experiment.  These experiments are expected to provide complementary sensitivity over a significant portion of the free parameter space presented in Fig.~\ref{Figs:Fig2}.  The milliQan experiment was proposed in 2016~\cite{Ball2016} and approved in the same year, supported largely by the sensitivity projections published in the preceding year~\cite{Haas2015} by several of the authors of the milliQan Letter of Intent.  The full milliQan detector system, located in the same gallery as their demonstrator detector, comprises two separate detector systems referred to as the ``bar'' and ``slab'' detectors~\cite{Ball2021}.  The bar detector is a significant upgrade over the demonstrator detector, containing approximately four times the surface area of active scintillator material and an additional layer of scintillator bars.  Specifically, the Run~3 bar detector comprises an array of $64$ plastic scintillator bars ($5 \times 5 \times 60$~cm each) divided equally among four longitudinal layers, with each bar coupled to a PMT~\cite{Ball2021}.  Additionally, the bar detector is surrounded by an active veto system of scintillator panels.  The slab detector is a separate detector system comprising four layers of scintillator slabs ($48$ slabs in total), each with a size of $40 \times 60 \times 5$~cm~\cite{Ball2021}.  In particular, the addition of the slab detector enables an increase in the overall detector sensitivity to high-mass mCPs.  In combination, the milliQan detector system could exclude at the $95$\% CL mCPs with masses and effective charges ranging from~$0.1 \lesssim m_{\chi} \lesssim 45$~GeV and $\sim0.003$--$0.3\:e$ (depending on the mCP mass), respectively, at the LHC's current Run~3 (assuming an integrated luminosity of $300$~fb$^{-1}$)~\cite{Ball2021}. Sensitivity projections for the milliQan bar detector to mCPs arising in the mixed-phase scenario, including both single- and double-mCP events, have been reported in~\cite{Izaguirre2015}, demonstrating significant sensitivity to the unique double-mCP signature. To the best of our knowledge, the only published projections for this model are those for the milliQan experiment.

\subsubsection{The FORMOSA Demonstrator}
The FORward MicrOcharge SeArch (FORMOSA) demonstrator~\cite{Citron2025} was installed in February 2024 in the UJ12 cavern at the LHC, adjacent to the FASER experiment, approximately $580$~m downstream of the ATLAS interaction point. The demonstrator comprises a $2 \times 2 \times 4$ array of $5$~cm $\times$ $5$~cm $\times$ $100$~cm scintillator bars, equipped with front, back, and side veto panels to tag beam muons and suppress beam-related backgrounds. By operating in the far-forward region, the demonstrator benefits from the exceptionally high flux of forward-produced mCPs in LHC $pp$ collisions. Commissioned during 2024 Run~3 collisions, it has demonstrated stable data acquisition and effective background rejection, with early 2025 upgrades achieving full hermeticity. A discussion of the full FORMOSA experiment, including projected sensitivity estimates for the HL-LHC, is provided in Sec.~\ref{sec:FORMOSA-FLARE}.

\subsubsection{The MoEDAL-MAPP Experiment}
The Monopole and Exotics Detector at the LHC (MoEDAL) is a pioneering experiment~\cite{doi:10.1098/rsta.2019.0382}---the first dedicated search experiment at the LHC---targeting highly ionizing particle avatars of new physics that could be missed by the general-purpose ATLAS and CMS experiments~\cite{doi:10.1142/S0217751X14300506,PhysRevLett.123.021802,Hirsch2021,Altakach2022,Acharya2022,PhysRevLett.126.071801}. In December 2021, the first phase of MoEDAL's Apparatus for Penetrating Particles (MAPP-1), a new stand-alone downstream detector, was approved by the CERN Research Board~\cite{Pinfold2791293} to search for feebly interacting particles, including mCPs. MAPP-1 is installed in the UA83 gallery at $\sim100$~m from IP8 and at an angle of $\sim7^{\circ}$ relative to the beamline~\cite{pinfold2023moedalmapp}---a different pseudorapidity region than the milliQan and FORMOSA detectors---providing complementary coverage of the kinematic distribution of mCPs between experiments. The sensitive volume of the MAPP-1 detector consists of four collinear sections, each with a cross-sectional area of $\sim1$~m$^{2}$ and comprising $100$ ($10 \times 10 \times 75$~cm) plastic scintillator bars read out by a single PMT each and placed in a four-fold coincidence.  MAPP-1 is housed within a hermetic veto system of scintillator panels, surrounded by an outer flame shield. The LHC Experiments Committee (LHCC) is currently reviewing the addition of an auxiliary detector, referred to as the MAPP Outrigger Detector (MAPP OD)~\cite{Pinfold:2918254}, designed to provide complementary sensitivity to high-mass, intermediate-charge mCPs ($Q_{\chi} \gtrsim 0.01\: e$), enhancing the overall physics reach of the MAPP experiment. The MAPP OD, comprising ten installation subunits each containing eight plastic scintillator slabs ($60$~cm $\times$ $30$~cm $\times$ $5$~cm) arranged in four layers, is planned to be installed in a duct between MAPP-1 and the beamline. Detailed studies of the projected sensitivities of MAPP-1 and the MAPP OD to mCPs at the HL-LHC have recently been published in Refs.~\cite{Kalliokoski2024} and \cite{Kalliokoski2025}, respectively.

\begin{figure}[htb!]
\centering
\includegraphics[width = 7.6 cm]{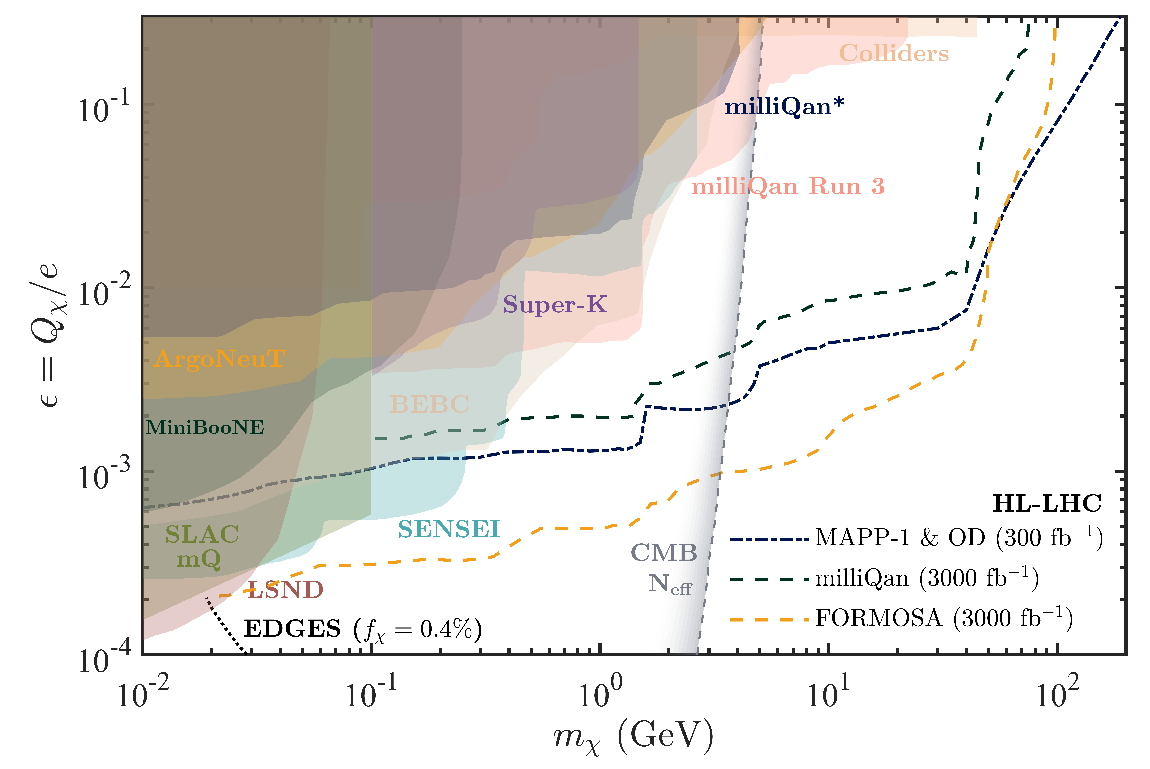}
\caption{Combined $95\%$~CL sensitivity projections for MAPP-1 and the MAPP OD to mCPs at the full HL-LHC ($\sqrt{s}=14$~TeV, $300$~fb$^{-1}$)~\cite{Kalliokoski2024,Kalliokoski2025}, compared with $95\%$~CL projections for FORMOSA and milliQan (combined slab and bar detectors) at the full HL-LHC ($\sqrt{s}=14$~TeV, $3000$~fb$^{-1}$)~\cite{Alemany2025}.}
\label{Figs:Fig5}
\end{figure}

Fig.~\ref{Figs:Fig5} presents the projected state-of-the-art sensitivities of the three main dedicated mCP search experiments at the HL-LHC: MAPP (MAPP-1 \& OD combined), milliQan (bar \& slab detectors combined), and FORMOSA~\cite{Alemany2025}. The three experiments are largely complementary in coverage, with FORMOSA achieving the greatest charge sensitivity, while the MAPP detector system achieves the highest reach in mCP mass, up to approximately $200$~GeV, due to the large geometric acceptance of the MAPP OD.

\subsection{Proposed Future Searches at Accelerators}\label{subsec:UpcomingSrch}
We conclude this section with a brief overview of the experiments proposed in the literature that could perform searches for mCPs in the near and far future.

\subsubsection{The \texorpdfstring{NA64$\mu$}{NA64μ} Experiment}
The NA64$\mu$ experiment~\cite{Gninenko:2653581}, which plans to operate at the high-intensity M2 muon beamline at the CERN SPS, has the primary aim of exploring dark sector models.  Using the same missing-energy technique as their proposed searches for dark sector mediator particles produced in the beam--target interactions, searches for mCPs can also be performed.  Gninenko et al. estimated the background-free sensitivity of the NA64$e$ and NA64$\mu$ experiments to mCPs, demonstrating the superior sensitivity of the latter experiment over the entire range of mCP parameter space covered~\cite{Gninenko2019}.  In particular, assuming a muon beam energy of $100$~GeV and a total muon flux of $N_{\mathrm{MOT}}=5\times10^{13}$, their analysis suggests that the NA64$\mu$ experiment could establish new limits (at the $90\%$ CL) for mCPs with masses ranging from approximately $0.1$ to $2.5$~GeV~\cite{Gninenko2019}.

\subsubsection{The FerMINI Experiment}
An mCP search at a proton fixed-target experiment---which could benefit from a larger mCP flux than collider experiments---has been proposed at Fermilab.  In particular, the aptly named Fermilab search for mCPs (FerMINI)~\cite{Kelly2019} has been proposed, which would deploy a MAPP/milliQan-like scintillation detector downstream from the existing NuMI beamline and/or the future DUNE beamline.  Considering one year of data collected from $120$~GeV protons on target at the NuMI beam $(N_{\mathrm{POT}}=6\times10^{20})$, sensitivity estimates suggest that, at the $95\%$ CL, FerMINI could probe a large portion of unconstrained parameter space covering mCP masses from approximately $0.03$ to $5$~GeV and reaching effective charges below $10^{-3}\:e$ over most of this range~\cite{Kelly2019}. Notably, due to the increased proton beam energy at the Fermilab NuMI beam compared to J-PARC, the production of mCPs via $\Upsilon$ decays is possible; thus, the FerMINI experiment could probe larger mCP masses than SUBMET.

\subsubsection{The SpinQuest Program}
While the FerMINI proposal has not advanced to an approved experimental stage, Fermilab has other proposed initiatives for mCP searches. The DarkQuest upgrade to the ongoing SpinQuest experiment has been proposed to extend the experiment's physics reach to dark-sector searches~\cite{Apyan2022}. Beyond this, a further proposed extension referred to as LongQuest, which carries forward certain ideas pioneered by FerMINI, considers the installation of additional detectors downstream of the SpinQuest apparatus in a low-background environment. In this configuration, a FerMINI-like scintillator detector could be deployed to search for mCPs produced at the target~\cite{Apyan2022}. Sensitivity projections for such a setup have been studied, considering a detector comprising a three-layer $18 \times 18$ scintillator-bar array located approximately $40$~m downstream of the target behind $10$~m of iron shielding. Assuming an exposure of $10^{20}$~POT at SpinQuest over $5$ years of operation, excellent sensitivity to mCPs in the mass range from $10^{-3}$ to $5$~GeV has been reported, reaching effective charges as low as $\sim 5 \times 10^{-5} \: e$~\cite{Bailloeul2025}.

\subsubsection{The FORMOSA and FLArE Experiments}\label{sec:FORMOSA-FLARE}
At the HL-LHC, the proposed Forward Physics Facility (FPF)---a new underground facility aimed at hosting four complementary far-forward detectors~\cite{ANCHORDOQUI2022,Anchordoqui2025}---has been discussed as a potential site for future dedicated search experiments for mCPs.  Among these, two FPF experiments with excellent sensitivity to mCPs have been proposed: the full FORMOSA experiment~\cite{Foroughi2021} and the Forward Liquid Argon Experiment (FLArE)~\cite{Kling2022}.  These experiments benefit from the enhanced production of mCPs in the forward direction exhibited by many of the principal modes of mCP production at hadron colliders. The FORMOSA detector design comprises a layered scintillation detector approximately four times larger than the MAPP-1 detector, though its final design will be constrained by the engineering limits of the FPF cavern. Based on the full detector design, FORMOSA's projected sensitivity at $13.6$~TeV, assuming an integrated luminosity of $2$~ab$^{-1}$, surpasses the charge sensitivities of the milliQan, MAPP-1, and FerMINI experiments~\cite{Anchordoqui2025}. An additional $\mathrm{Ce}\mathrm{Br}_{3}$ subdetector has also been considered, demonstrating further improvements in sensitivity for charges below $\sim 10^{-3} \:e$. In comparison, the proposed FLArE detector, located further from the IP than FORMOSA, is projected to achieve a sensitivity complementary to the MAPP-1 and milliQan experiments~\cite{Kling2022}.

\subsubsection{The SHiP Experiment}
The Search for Hidden Particles (SHiP)~\cite{Ahdida2019} is a proposed general-purpose experiment aimed at searches for dark sector particles at the intensity frontier. SHiP will be hosted at the ECN3 facility, using a $400$~GeV proton beam from the CERN SPS. Among the numerous new physics scenarios that comprise the thorough physics program of the SHiP experiment~\cite{Alekhin2016}, stringent searches for mCPs with masses at the MeV--GeV scale are possible.  The projected sensitivity of the SHiP experiment to mCPs was estimated in Ref.~\cite{Magill2019}, which assumed a sample of $N_{\mathrm{POT}} = 2 \times 10^{20}$ and a distance of $50$~m from the target to the detector.  Excellent sensitivity to mCPs was reported, which could exceed current bounds for mCPs with masses from approximately $0.1$ to $5$~GeV by roughly an order of magnitude, and would provide complementary limits to the projections for the MAPP-1, milliQan, and FORMOSA experiments at the HL-LHC. Several additions to SHiP offering significant enhancements to its projected sensitivity to mCPs have been proposed: (1) Ref.~\cite{Bailloeul2025}, which considers a detector comprising an array of scintillator bars, and (2) Ref.~\cite{Ferrillo2024}, which considers an LArTPC setup similar to ArgoNeuT.

\subsubsection{The LANSCE-mQ Experiment}
The LANSCE-mQ experiment~\cite{Tsai2026} is a proposed search for mCPs using an $800$~MeV proton beam incident on a fixed target at the Los Alamos Neutron Science Center (LANSCE). The detector design consists of two layers of segmented scintillation bars, implemented with either plastic (e.g., EJ-200) or high-light-yield $\mathrm{Ce}\mathrm{Br}_{3}$, coupled to PMTs to detect the anomalously low ionization signals produced by mCPs. Two candidate detector locations have been identified, at approximately $6$~m (ER1) and $35$~m (ER2) from the target; the former represents the more ambitious configuration, since operation at ER1 is contingent upon achieving sufficient suppression of beam-induced neutron backgrounds. \textsc{Geant}4 simulations and in situ measurements indicate that backgrounds from beam-induced neutrons, cosmic rays, and photodetector dark counts can be mitigated through shielding and timing-based coincidence requirements. For nominal operation over six years at either location, collecting $N_{\mathrm{POT}}= 5.9 \times 10^{22}$, the LANSCE-mQ experiment is projected to achieve leading $95\%$ CL sensitivity to mCPs with masses in the range $\sim 1$--$300$~MeV~\cite{Tsai2026}.

\section{Conclusions}\label{sec:Conc}
After over $40$ years of exhaustive searches, no mCPs have yet been discovered.  Notwithstanding these efforts, numerous scenarios involving mCPs remain viable and interesting (given their potential to provide insights into various mysteries, such as the nature of dark matter and electric charge quantization); the search thus continues.  In particular, a substantial window of unconstrained mCP parameter space is currently accessible to contemporary accelerator facilities.  Several dedicated search experiments targeting mCPs produced in high-energy $pp$ collisions at the LHC have received approval and are expected to achieve unprecedented sensitivity to mCPs with masses at the GeV scale. For these experiments---MoEDAL-MAPP, milliQan, and FORMOSA---we have summarized the projected $95\%$ confidence level sensitivity estimates for the HL-LHC. The projections highlight the state of the art in accelerator searches for mCPs, demonstrating complementary reach, with FORMOSA achieving the greatest charge sensitivity, and the combined MAPP-1 and MAPP OD system offering leading sensitivity to high-mass mCPs, reaching masses up to approximately $200$~GeV. Collectively, these efforts promise substantial advances in accelerator searches for mCPs over the next decade.


\backmatter

\bmhead{Acknowledgments}
We are grateful to the Natural Sciences and Engineering Research Council of Canada (NSERC) for partial financial support: Discovery Grant, SAPPJ-2019-00040; Research Tools and Instruments Grants, SAPEQ-2020-00001 and SAPEQ-2022-00005. M.d.M. thanks NSERC for partial financial support under Discovery Grant No. RGPIN-2016-04309. M.S. acknowledges support by the Generalitat Valenciana via the APOSTD Grant No. CIAPOS/2021/88 and the Excellence Grant No. CIPROM/2021/073, as well as by the Spanish MCIN/AEI/10.13039/501100011033/ and the European Union/FEDER via the Grant PID2021-122134NB-C21. 

\section*{Statements and Declarations}

\bmhead{Data Availability Statement}
The results reported in this study were obtained from analyses of simulation data; no original datasets were created.

\bmhead{Competing Interests}
The authors declare no competing or conflicting interests.

\bmhead{Author Contributions}
The study was conceptualized by M.S. The original draft of the manuscript was written by M.S. and reviewed and edited by M.d.M., P.-P.A.O., J.P., A.S., and M.S.  The figures were produced by M.S.  Project supervision was provided by J.P.  All authors have read and agreed to the published version of the manuscript. 



\bibliography{Minicharge}


\end{document}